**Suppression of long-range ordering and multiferroicity in Sr-substituted $Ba_{3-x}Sr_xMnNb_2O_9$ (x = 1 and 3)**

Shivani Sharma[1,2*], Premakumar Yanda[1], Poonam Yadav[3], Ivan da Silva[2], A. Sundaresan[1]

[1]*School of Advanced Materials and Chemistry and Physics of Materials Unit, Jawaharlal Nehru Center for Advanced Scientific Research, Jakkur P.O., Bangalore 560064, India*

[2]*ISIS Facility, Rutherford Appleton Laboratory, Chilton, Didcot OX11 0QX, United Kingdom*

[3]*UGC-DAE Consortium for Scientific Research, Indore 452001, India*

## Abstract

Effects of Sr substitution at A-site in ordered perovskite $Ba_{3-x}Sr_xMnNb_2O_9$ (x = 1 and 3) have been investigated using X-ray diffraction, magnetization, dielectric/magnetodielectric and neutron diffraction measurements. The parent compound $Ba_3MnNb_2O_9$ having a large spin (S=5/2) is known to exhibit type-II multiferroic properties with quasi 2D triangular lattice antiferromagnetic ground state. A slight perturbation in exchange interaction due to substitution of smaller size isovalent ion at the A-site in $Ba_{3-x}Sr_xMnNb_2O_9$ (x = 1 and 3) has been found to alter the ground states drastically and hence the multiferroicity. The crucial role of various fluctuations (quantum and/or thermal), weak lattice distortion induced by Sr-substitution and slight imbalance between different fluctuations in determining the ground states and the multiferroicity is discussed and compared with the results of smaller spin compounds (S = 1/2 or 1).

***sharma@magnet.fsu.edu**

## Introduction

Two-dimensional (2D) triangular lattice antiferromagnetic (TLAF) state can be easily realized in a specific type of B-site ordered (1:2) perovskites having the general formula $A_3BB'_2O_9$ where the magnetic B-cations form the TLAF structure [1-8]. For most of these 2D TLAFs, the magnetic ground state comprises of highly degenerate 120° orientated spin structure in zero field whereas, in the presence of a certain magnetic field, the structure deviates from 120° spin configuration resulting to up-up-down (*uud*) phase [1-6]. In *uud* phase, the magnetization exhibits a plateau with 1/3 value of the saturation magnetization, similar to typical examples showing this state [9-12]. A third magnetic phase has also been realized for few systems even at higher magnetic fields [1,3]. The driving mechanism behind the exotic ground states of these compounds can be understood based on recent results in several other members of this family [1-6]. These studies in addition to geometrical frustration arising from TLAF arrangement, suggest that quantum and thermal fluctuations also play an important role in governing the exotic ground states. Since the effect of quantum fluctuation is more pronounced for small spin systems (S = 1/2 or 1) as compared to large spin number (S = 5/2) where the thermal fluctuations play the dominating role in governing the ground state [1-6], the role of spin number in deciding the effect of fluctuations on *uud* phase is unavoidable. Since the magnetic ground states depend on the spin number, the multiferroic behaviour observed for these systems also has relevance with the spin number. For example, in small spin systems (S = 1/2 or 1), the quantum fluctuations establish the magnetic ground states and the multiferroicity is achieved in all the experimentally observed magnetic ordered phase [1,3,5,6]. For example, $Ba_3NiNb_2O_9$ demonstrates multiferroic properties in all the three different magnetic phases reported for this system and the ferroelectricity appears at each phase boundary, making it a unique example showing the most possible magnetic phases [3]. However, the large spin system $Ba_3MnNb_2O_9$ exhibits a suppressed *uud* phase due to the dominating thermal fluctuation mechanism and the multiferroicity is realized only in the non-chiral magnetic state [4].

Apart from the quantum and thermal fluctuations, another important factor that is expected to play a significant role in deciding the ground state is the lattice distortion. The effect of weak lattice distortion has been recently studied in small spin compound $Sr_3NiNb_2O_9$ where the replacement of Ba-ions by smaller Sr-ion leads to the transformation of equilateral TLAF to isosceles plaquette with one longer (weak) and two shorter (strong) bonds [5]. These results suggest that the weak lattice distortion doesn't alter the magnetic ground state or multiferroic character of small spin systems like $Sr_3NiNb_2O_9$ and $Sr_3NiTa_2O_9$ [5,13]. For these, the ferroelectric phase transition and the magnetic phase transition occur simultaneously at $T_{N2}$. However, the imbalanced exchange interactions induced by the lattice distortion in $Sr_3NiNb_2O_9$ change the easy plane anisotropy of the Ba compound to the easy-axis anisotropy in the Sr compound [5]. Also, the *uud* phase is stabilized in a narrower magnetic field region in the Sr compound, possibly due to the reduced quantum fluctuations in the isosceles triangular lattice.

In the present report, we show that a weak lattice distortion in a large spin system is enough to suppress the long range ordered magnetic ground state and the multiferroic properties. We have studied Sr substituted $Ba_{3-x}Sr_xMnNb_2O_9$ (x = 1 and 3) compounds with S=5/2 to understand the role of weak lattice distortion in deciding the magnetic and multiferroic properties. The parent compound $Ba_3MnNb_2O_9$ shows a narrow two-step transition at $T_{N1}$ = 3.4 K and $T_{N2}$ = 3.0 K [4]. For $Ba_3MnNb_2O_9$, the neutron diffraction measurements supported by DFT calculation indicate a 120◦ spin structure in the *ab* plane with out-of-plane canting at low temperatures which evolves into up-up-down (*uud*) and oblique phases showing successive magnetic phase transitions with increasing magnetic field [4]. Intriguingly, the multiferroicity is observed only when the spins are not collinear but suppressed in the *uud* and oblique phases [4]. Our x-ray diffraction, magnetic, dielectric/magnetodielectric and neutron diffraction results on Sr-doped compounds show that the weak lattice distortion induced by Sr substitution is sufficient to destroy the long-range ordered phase and multiferroicity in this large spin compounds, contrary to small spin compounds [5,6]. The role of various fluctuations (quantum and/or thermal) and weak lattice distortion induced by Sr substitution plays a dominating role in determining the ground states and the multiferroicity for these large spin (S = 5/2) compounds which is discussed in detail by comparing the results with small spin compounds (S = 1/2 or 1).

**Experimental details**

The polycrystalline samples of $Ba_{3-x}Sr_xMnNb_2O_9$ (x = 0, 1 & 3) were prepared through a solid-state reaction route using a stoichiometric mixture of $BaCO_3$, $SrCO_3$, $MnO_2$ and $Nb_2O_5$, following the procedure described in Ref. 4. Powder X-ray diffraction (XRD) data in the $2\theta$ range of 10-100° were recorded using the PANalytical Empyrean diffractometer with Cu-$K_\alpha$ radiation to confirm the phase formation. The temperature, field and time dependent *dc*-magnetization measurements were carried out using SQUID MPMS3 under various protocols like zero field cooled (ZFC) and field cooled (FC) conditions, with cooling and heating rates of 3 K/min. The temperature-dependent neutron diffraction experiments were carried out using GEM instrument at the ISIS Neutron and Muon Source, UK. Magnetodielectric (MD) measurements have been done using Agilent E4980A LCR meter and temperature and magnetic field control provided by Quantum Design's PPMS. Rietveld analysis of X-ray and neutron diffraction data has been performed using JANA2006 [14].

**Results and Discussion**

Figure 1(a-b) represents the Rietveld refined room temperature XRD patterns of $Ba_{3-x}Sr_xMnNb_2O_9$ (x = 1 & 3). Partial substitution of Ba-ion by Sr-ion in $Ba_2SrMnNb_2O_9$ doesn't alter the lattice symmetry and leads to the same space group as that of the parent compound $Ba_3MnNb_2O_9$. Therefore, the XRD pattern of $Ba_2SrMnNb_2O_9$ has been refined using trigonal space group *P*-3*m*1, consistent with the parent compound[4]. The complete substitution of Ba-ion by Sr-ion in $Sr_3MnNb_2O_9$ induces considerable lattice distortion resulting in the monoclinic lattice. The $Sr_3MnNb_2O_9$ is refined using monoclinic space group $P2_1/c$ with the transformation of equilateral Mn-triangle into isosceles triangle. Here, two longer (weak) bond lengths of 5.6568(3) Å and one shorter (strong) bond length of 5.6113(6) Å have been observed, contrary to smaller spin (S = 1) compound $Sr_3NiNb_2O_9$ where two strong (shorter) and one weak (longer) bond was reported [5]. It is worth mentioning here that the structure of $Ba_2SrMnNb_2O_9$ consists of equilateral Mn-triangles having a bond length of 5.7768(5) Å. The refined lattice parameters for $Ba_2SrMnNb_2O_9$ are $a$ = 5.7729(1) Å and $c$ = 7.0725(2) Å where $\gamma$ = 120° whereas for $Sr_3MnNb_2O_9$ are $a$ = 9.8283(3) Å, $b$ = 5.6105(2) Å, $c$ = 17.0435(2) Å and $\beta$ = 125.344(1) °. The monoclinic distortion in $Sr_3MnNb_2O_9$ is comparable with that of $Sr_3NiNb_2O_9$ [5]. The detailed structural parameters including Wyckoff positions alongside the reliability factors are given in table I for $Sr_3MnNb_2O_9$. A small amount of $Sr_2Nb_2O_7$ non-magnetic impurity phase is also found in $Sr_3MnNb_2O_9$ which is indicated by the asterisks in Fig. 1b.

Figure 2(a,b) represents the *dc* magnetization behaviour of $Ba_2SrMnNb_2O_9$ and $Sr_3MnNb_2O_9$ as a function of temperature measured under an applied magnetic field of 100 Oe in zero field cooled (ZFC) and field cooled (FC) conditions. Both the compounds exhibit a peak at low temperature that is denoted by $T_f$ which indicates the frozen or spin glass transition temperature of Mn spins. The justification to call this as frozen or spin-glass transition temperature will be given later. The value of $T_f$ for $Ba_2SrMnNb_2O_9$ and $Sr_3MnNb_2O_9$ is ~ 3 and 6 K, respectively which is followed by a bifurcation between ZFC and FC curves below $T_f$. The upper insets in Fig. 2(a,b) represents the $\chi_{dc}$ and the inverse susceptibility over a broad temperature range of 2-400 K. The Curie-Weiss (CW) fit to the inverse susceptibility data leads to the value of effective paramagnetic moments to be 5.94 and 6.09 $\mu_B$ which is consistent with the spin only value for $Mn^{2+}$ spins (5.92 $\mu_B$). The value of Weiss constant ($\theta_c$) is -4.15(1) and -6.74(3) K for $Ba_2SrMnNb_2O_9$ and $Sr_3MnNb_2O_9$, respectively. The negative value of $\theta_c$ further indicates the antiferromagnetic (AFM) nature of magnetic interactions and slightly large value of $\theta_c$ for $Sr_3MnNb_2O_9$ indicates that the exchange coupling increases with Sr substitution. The lower insets in Fig. 2(a,b) shows the isothermal magnetization curves measured at 300 and 2 K for respective samples. The 300 K curve is typically linear as expected from a paramagnetic state and 2 K curve exhibits a S-shape curve, similar to one observed for many other spin-glass systems [15,16,17]. The overall $\chi_{dc}$ response of $Ba_2SrMnNb_2O_9$ is looks identical to the parent compound $Ba_3MnNb_2O_9$ [4] except the transition temperature whereas the response of $Sr_3MnNb_2O_9$ is almost matching with small spin (S = 1) $Sr_3NiNb_2O_9$ compound [5].

To study the dynamic response of the anomalies observed in *dc* magnetic susceptibility data, *ac* susceptibility ($\chi_{ac}$) has been measured as a function of temperature at different frequencies. Figure 3(a,b) represents the $\chi_{ac}$ behaviour of both compounds as a function of temperature over a broad range of frequency extending from 10 Hz to 10 kHz. A clear dispersion and shift in peak position have been observed for both the compositions which indicates the possibility of spin glass ordering at $T_f$. This result contrasts with the other members of this family like $Ba_3MnNb_2O_9$, $Sr_3NiNb_2O_9$, $Ba_3CoSb_2O_9$, $Ba_3NiNb_2O_9$, $Ba_3NiNb_2O_9$ etc. which are reported to exhibit long-range ordered magnetic ground states [1-8].

To explore the possible relaxation in the case of spin glass ordering, the thermoremanent magnetization of $Sr_3MnNb_2O_9$ has been measured in ZFC and FC conditions to characterize the relaxation behaviour in the vicinity of $T_f$. For ZFC curves, the sample was cooled from 300 K (PM region) down to 2 K and then 500 Oe field was applied to record the magnetization $M(t)$ as a function of time (t). For FC $M(t)$ curves, the sample was cooled in the presence of field H = 200 Oe down to 2 K and $M(t)$ was recorded as a function of time after removing the field. The ZFC and FC M($t$) curves are presented in Fig. 4. The red line in Fig. 4 is the stretched exponential fit of the data [18]. The fitting parameters $\tau$ and $\beta$ are the characteristic time constant and critical exponent respectively and the reported value of $\beta$ lies between 0 and 1 [17,19,20,21] for spin glass systems. The value of $\tau$ and $\beta$ for ZFC and FC curves are mentioned in respective figures. The relaxation mechanism at 2 K is very slow, and the value of $\tau$ approaches 10 ks under ZFC. Therefore, our detailed relaxation results confirm the spin glass state in $Sr_3MnNb_2O_9$ below $T_f$. Due to the experimental temperature limit, we could not measure the thermoremanent behaviour of $Ba_2SrMnNb_2O_9$ but in the light of other results and similarities with $Sr_3MnNb_2O_9$, we suggest that the ground state is spin glass in this composition as well.

To firmly establish the magnetic ground state in both compositions, neutron powder diffraction (NPD) data have been collected at 100 and 2 K. Figure 5 (a-b) represents the Rietveld refined NPD plot of $Ba_2SrMnNb_2O_9$ and $Sr_3MnNb_2O_9$ respectively recorded at 2 K. The insets in respective figures represent the stack of 100 and 2 K NPD patterns, respectively. Furthermore, no new peak or significant change in the background below the magnetic anomaly is evident for studied systems. This confirms that the anomalies observed in *dc* and *ac* magnetization data near 3 and 6 K are related to the spin glass ordering of Mn moments in these systems. Also, there is no appreciable change in peak profile with temperature which confirms that the structures remain the same down to 2 K.

Further, to investigate the magnetodielectric (MD) response of the as prepared samples, dielectric constant and loss has been measured at 50 Hz frequency under various applied magnetic field ranging from 0-8 Tesla (T) and presented in Fig. 6(a-b). The zero-field data exhibits a sharp drop in dielectric permittivity ($\varepsilon'$) values below $T_f$ for both compounds, suggesting the presence of MD coupling. The effect of the magnetic field is clear by the suppression of this drop with increasing applied field for both the compounds. With an applied field of 8 T, the drop almost vanishes, and no clear anomaly exists at this field. The loss curve also shows the effect of the applied field on dielectric behaviour. For $Ba_2SrMnNb_2O_9$, the loss exhibits a small upturn below $T_f$ whereas for $Sr_3MnNb_2O_9$, it shows a peak at $T_f$. As a result of the applied field, both the upturn and peak suppress with increasing field. Concomitantly to confirm and to investigate the detailed MD response, the isothermal MD behaviour is studied at three temperatures in the vicinity of $T_f$ (above and below) for both compositions. Fig. 6(c-d) shows the MD ($\Delta\varepsilon', \%$) response of $Ba_2SrMnNb_2O_9$ and $Sr_3MnNb_2O_9$, respectively. A positive MD effect at 2 K is observed for both compounds and the typical values of MD is ~ 0.055 and 0.27 % at 2 K and 8 T for $Ba_2SrMnNb_2O_9$ and $Sr_3MnNb_2O_9$, respectively. The higher value of MD for $Sr_3MnNb_2O_9$ can be attributed to the enhanced lattice distortion induced by the complete substitution of Ba-ion by smaller Sr-ion. Further, to check the multiferroic/magnetoelectric response of these systems, pyrocurrent measurements were also performed but we didn't get any pyro peak or anomaly in the vicinity of $T_f$ within our experimental resolution limit.

The disappearance of long-range ordering upon substitution of Ba atom by Sr in $Ba_2SrMnNb_2O_9$ indicates the important role of the size of A-site cation in deciding magnetic and electric ground states and hence the multiferroic behaviour of these systems. However, the existence of multiferroicity and long-range ordering in $Sr_3NiNb_2O_9$ [5] indicates the importance of the co-operative role of A and B-site cation in deciding the ground state of these partially ordered perovskites. The substitution of smaller size Sr cations is inducing microstructural strain effects which at first seem to govern the multiferroic properties. However, the occurrence of multiferroicity in $Sr_3NiNb_2O_9$ raises the doubt about the sole responsibility of Sr-induced strain to destroy the multiferroic behaviour. Lee *et al.* [5] have studied that the equilateral Mn triangle observed for $Ba_3MnNb_2O_9$ transforms to isosceles Ni- triangle for $Sr_3NiNb_2O_9$. The same situation exists for $Sr_3MnNb_2O_9$ whereas for partially substituted Ba compound i.e. $Ba_2SrMnNb_2O_9$, the equilateral triangle and the lattice symmetry remain intact with Sr substitution. We suggest that the possible mechanism behind the observed behaviour is the result of weak lattice distortion induced with Sr substitution which leads to the inequivalent TLAF arrangement. The combined effect of large spin number and inequivalent TLAF interactions leads to the reduction in quantum fluctuations which are one of the important factors in governing the magnetic ground state in these systems. In the absence of these fluctuations, the dominating thermal fluctuations lead to a short-range ordered state in studied compounds. Therefore, it appears that a slight change or imbalance in magnetic exchange interaction induced through structural distortion can completely change the magnetic and electric ground state and hence the multiferroic behaviour in this family. This study confirms that not only the size of both A and B-site cations but also the spin number has a significant role in deciding the multiferroic behaviour of these systems. Suppression of multiferroicity is also reported for 10 % Al-doped $TbMnO_3$ whereas the long-range magnetic ground state was still observed [22]. Recently, Upadhyay et al. [23] have observed the destruction of multiferroicity in $Tb_2BaNiO_5$ with even 10 % substitution of Sr atoms at Ba-site with the drastic reduction in magnetodielectric effect with respect to the parent compound.

## Conclusion:

In summary, $Ba_2SrMnNb_2O_9$ and $Sr_3MnNb_2O_9$ are synthesized successfully and the refined XRD data confirms the trigonal structure for $Ba_2SrMnNb_2O_9$ and monoclinic for $Sr_3MnNb_2O_9$. Despite being isostructural with $Ba_3MnNb_2O_9$ (as $Ba_2SrMnNb_2O_9$) and $Sr_3NiNb_2O_9$ (as $Sr_3MnNb_2O_9$) and exhibiting strikingly similarities between bulk *dc* magnetization behaviour, no long-range ordered magnetic ground state has been observed. The thermoremanent relaxation and *ac* susceptibility data indicate the spin glass transition which is confirmed via temperature dependent neutron diffraction results. $Ba_2SrMnNb_2O_9$ and $Sr_3MnNb_2O_9$ undergo spin glass state below ~ 3 and 6 K, respectively. The magnetodielectric effect is observed for both compounds whereas the absence of ferroelectricity is confirmed via pyrocurrent measurements. This is worth mentioning here that most of the other members of this family like $Ba_3CoNb_2O_9$, $Ba_3MnNb_2O_9$, $Sr_3CoNb_2O_9$, $Sr_3NiNb_2O_9$, etc. are known to exhibit multiferroic behaviour particularly at low temperatures [1-8].

## Acknowledgement:

Authors would like to acknowledge Sheikh Saqr Laboratory (SSL) and International Centre for Materials Science (ICMS) at Jawaharlal Nehru Centre for Advanced Scientific Research (JNCASR) for various experimental facilities. Shivani Sharma would like to thank India-Nanomission, DST for funding the postdoctoral position. The authors thank the ISIS facility for beam time for neutron measurements. P.Y acknowledges the University Grants Commission (UGC) for Ph.D. Fellowship (Award No. 2121450729).

**Table I:** Refined Wyckoff positions for $Sr_3MnNb_2O_9$. Reliability factors are: Rp = 6.79, Rwp. = 9.70, and GoF = 4.29.

| Atom | Site | *x* | *y* | *z* | Occupancies |
|---|---|---|---|---|---|
| Sr1 | 4e | 0.1541(7) | 0.5307(0) | 0.0931(2) | 1.000 |
| Sr2 | 4e | 1.0232(8) | 0.0413(6) | 0.0958(2) | 1.000 |
| Sr3 | 4e | 0.6695(8) | -0.3950(3) | 0.3463(6) | 1.000 |
| Mn1 | 2a | 0.0000 | 0.0000 | 0.0000 | 0.500 |
| Mn2 | 2d | 0.5000 | 0.5000 | 0.0000 | 0.500 |
| Nb1 | 4e | 0.5525(5) | 0.4944(8) | 0.3613(8) | 1.000 |
| Nb2 | 4e | 0.0962(3) | -0.0511(0) | 0.3904(8) | 1.000 |
| O1 | 4e | 0.9825(8) | 0.6016(4) | 0.2457(8) | 1.000 |
| O2 | 4e | 0.5382(5) | 0.8800(2) | 0.2809(4) | 1.000 |
| O3 | 4e | 0.2063(6) | 0.4098(0) | 0.2404(5) | 1.000 |
| O4 | 4e | 0.9850(7) | 0.6150(8) | 0.9092(4) | 1.000 |
| O5 | 4e | 0.0352(2) | 0.4170(2) | 0.9292(1) | 1.000 |
| O6 | 4e | 0.4661(9) | 0.2651(8) | 0.8848(7) | 1.000 |
| O7 | 4e | 0.4967(2) | 0.8766(2) | 0.9187(5) | 1.000 |
| O8 | 4e | 0.7315(2) | 0.0271(9) | 0.9030(6) | 1.000 |
| O9 | 4e | 0.2536(2) | 0.5965(3) | 0.9127(8) | 1.000 |

**Figure Captions:**

**Fig. 1:** Rietveld refined room temperature XRD patterns of (a) $Ba_2SrMnNb_2O_9$ using trigonal space group $P$-3$m$1 and (b) $Sr_3MnNb_2O_9$ using monoclinic symmetry $P2_1/c$.

**Fig. 2:** The temperature-dependent *dc* susceptibility ($\chi_{dc}$) behaviour of (a) $Ba_2SrMnNb_2O_9$ and (b) $Sr_3MnNb_2O_9$ measured with 100 Oe applied field under zero field (ZFC) and field cooled (FC) conditions. The upper insets in the respective figures show the $\chi_{dc}$-*vs*-*T* behaviour over a broad temperature range starting from 2-400 K. The inverse susceptibility is also plotted in the same inset with the linked *x*-axis and right *y*-axis. The lower insets in (a) and (b) exhibit the isothermal magnetization behaviour measured at 2 and 300 K.

**Fig. 3:** Temperature-dependent *ac* magnetization ($M'$) behaviour of (a) $Ba_2SrMnNb_2O_9$ and (b) $Sr_3MnNb_2O_9$ with a frequency ranging from 10 Hz to 10 kHz.

**Fig. 4:** Relaxation of ZFC and FC magnetization as a function of time at 2 K. The solid red line indicates the stretched exponential fit of the experimental data. The values of relaxation time ($\tau$) and critical exponent ($\beta$) are mentioned.

**Fig. 5:** Rietveld refined neutron powder diffraction pattern of (a) $Ba_2SrMnNb_2O_9$ and (b) $Sr_3MnNb_2O_9$ recorded at 2 K. The used space groups are mentioned in respective figures. The insets in the respective figure show the stacks of 2 and 100 K pattern for corresponding compounds. The goodness of fit parameters for $Ba_2SrMnNb_2O_9$ are: GoF = 2.16, $R_{wp}$ = 3.98 and for $Sr_3MnNb_2O_9$, the values are GoF = 1.35 and $R_{wp}$ = 3.10.

**Fig. 6:** Temperature-dependent dielectric permittivity ($\varepsilon'$) response of (a) $Ba_2SrMnNb_2O_9$ and (b) $Sr_3MnNb_2O_9$ measured with 50 kHz frequency under various applied magnetic fields ranging from 0 to 8 T. The insets in respective figures shows corresponding loss spectra. The curves in (c) and (d) shows the isothermal magnetodielectric ($\Delta\varepsilon'$, % ) response of $Ba_2SrMnNb_2O_9$ and $Sr_3MnNb_2O_9$ respectively.

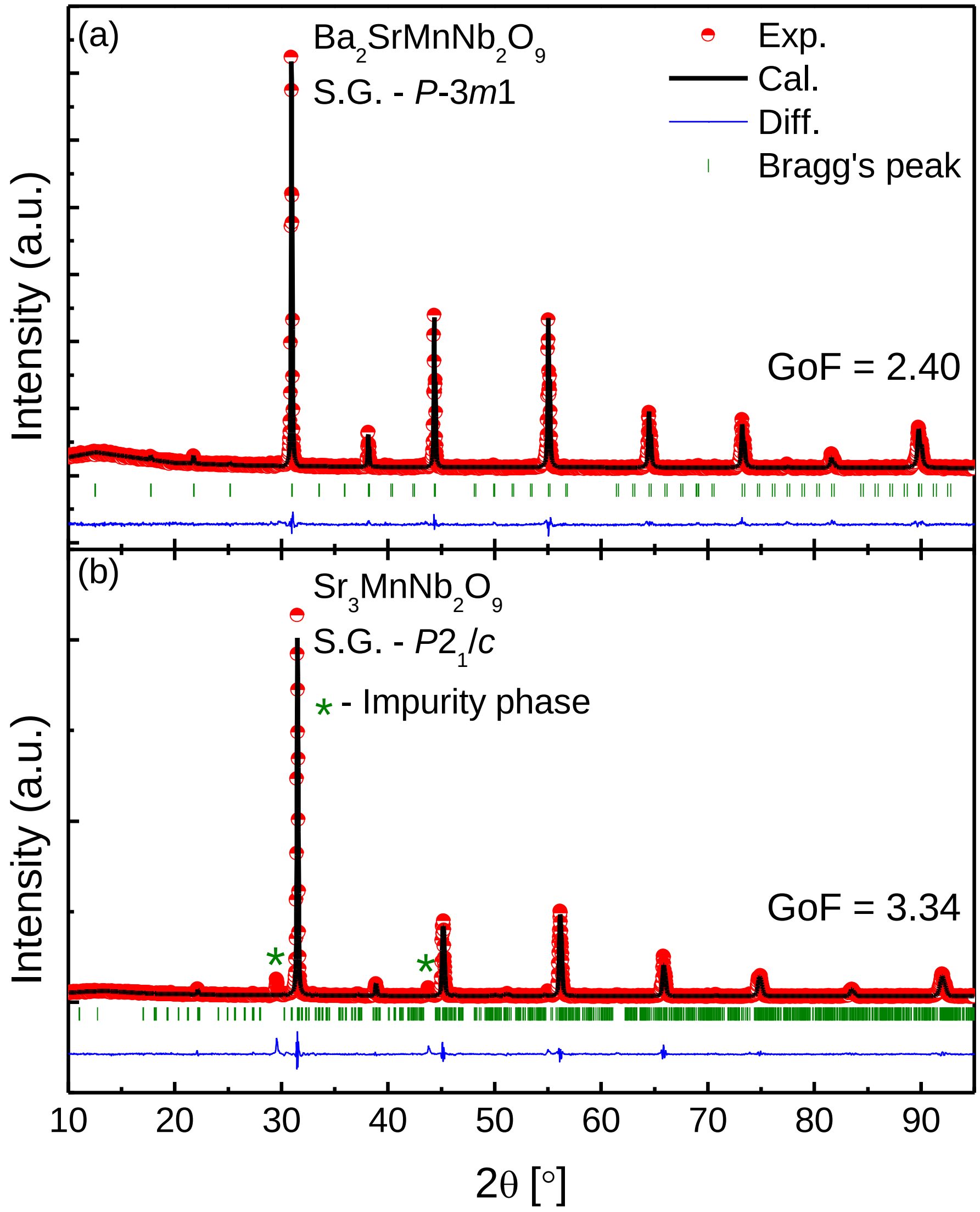

Figure 1

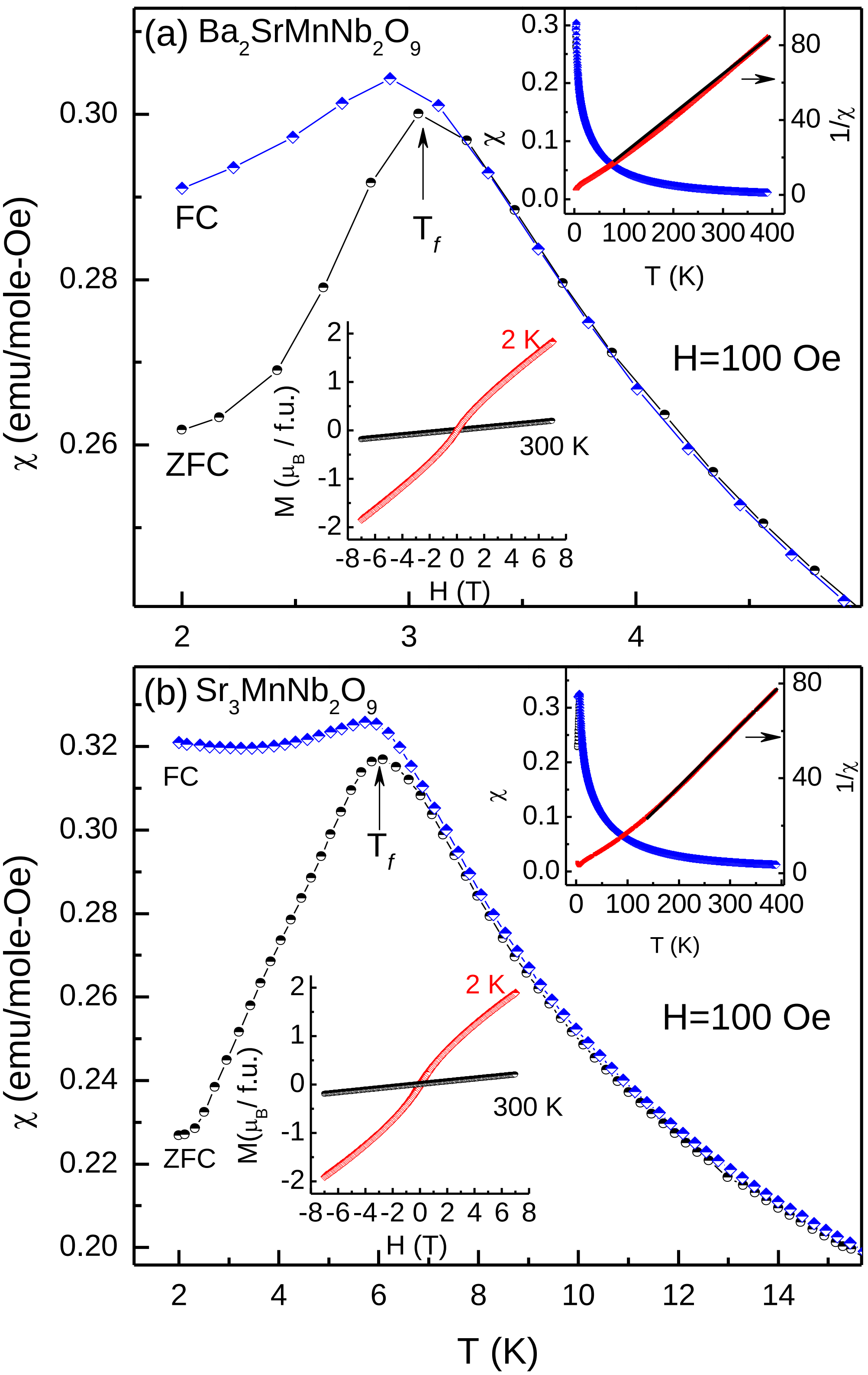

Figure 2

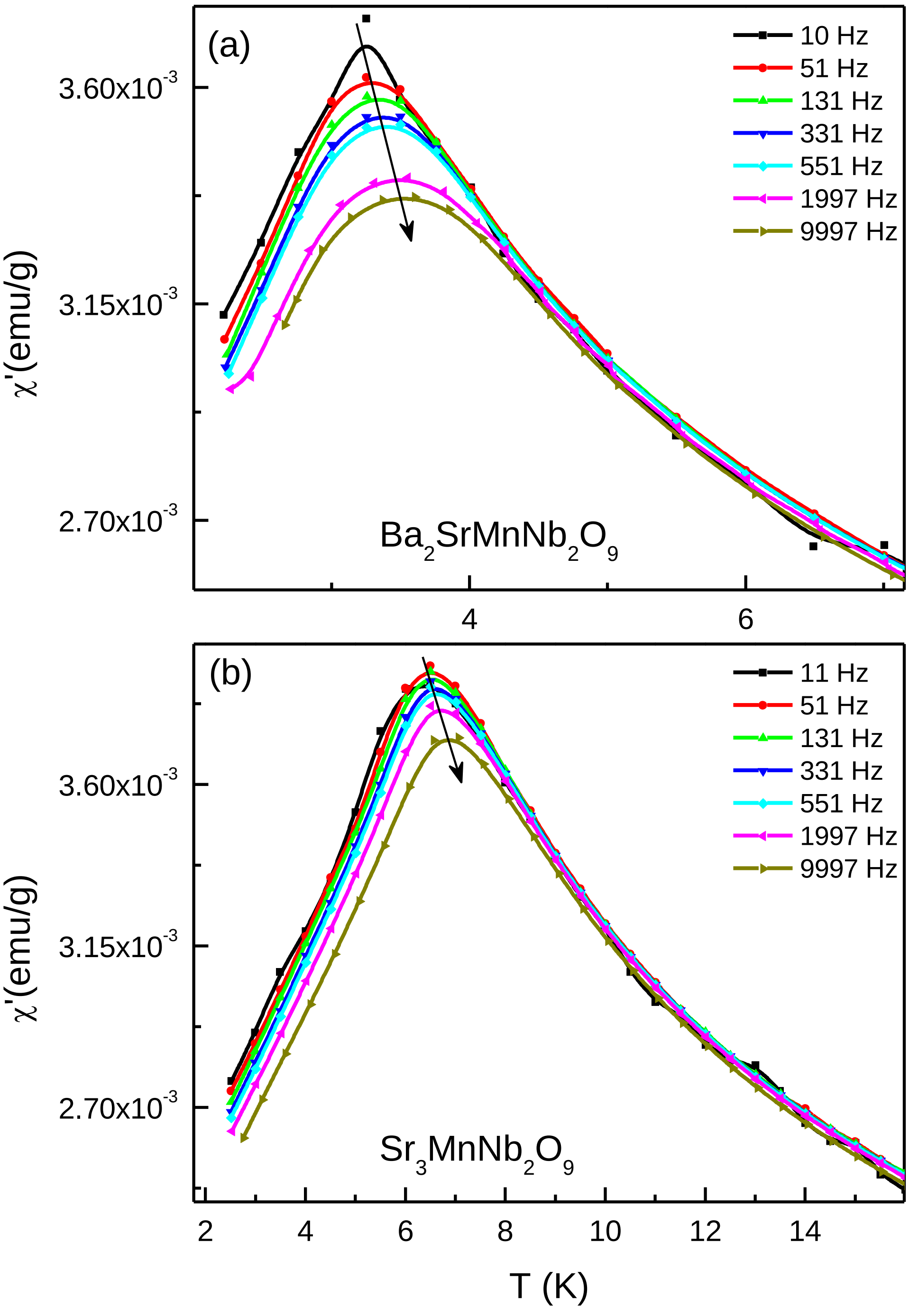

Figure 3

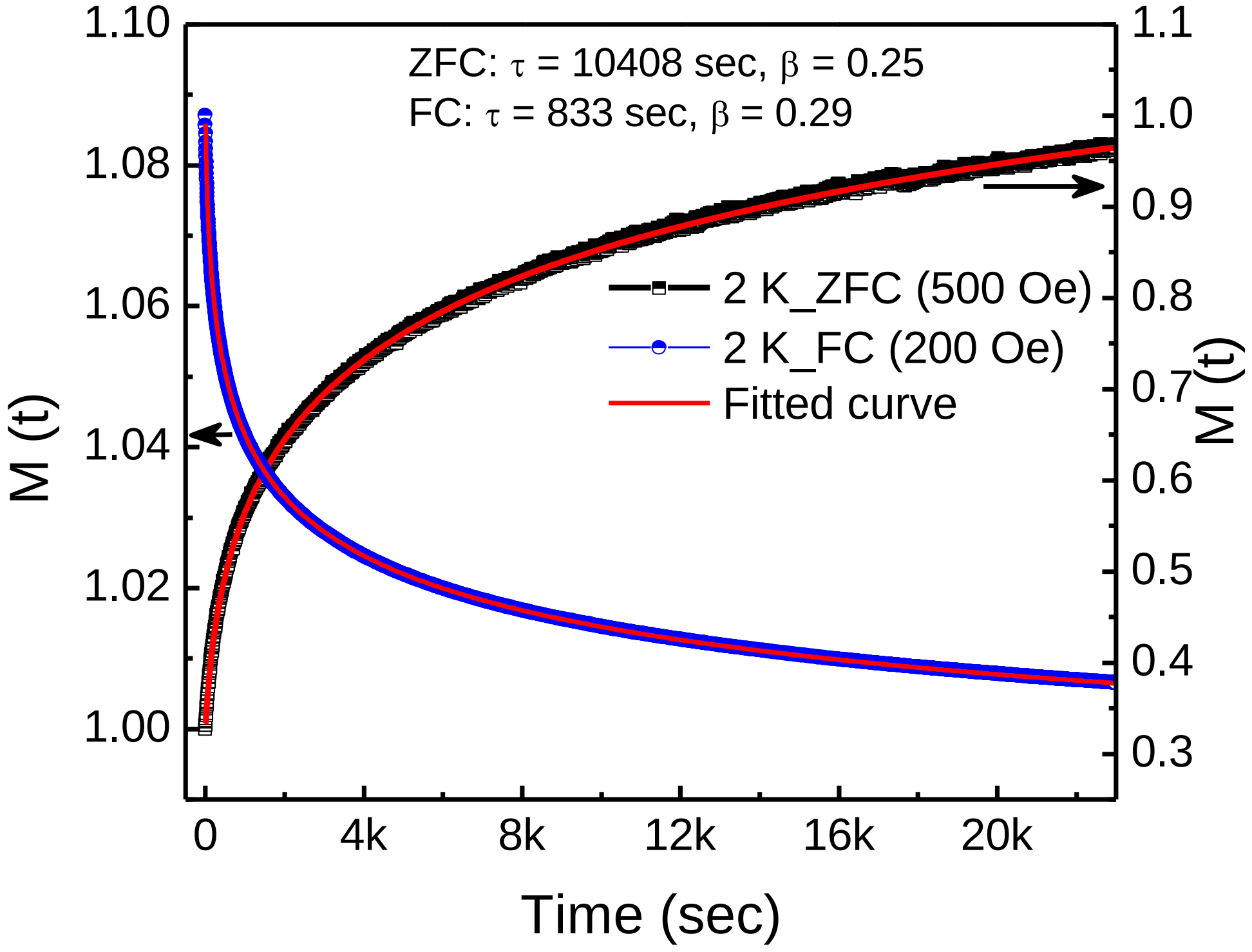
ZFC: τ = 10408 sec, β = 0.25
FC: τ = 833 sec, β = 0.29
2 K_ZFC (500 Oe)
2 K_FC (200 Oe)
Fitted curve
M (t)
M (t)
Time (sec)
1.10
1.08
1.06
1.04
1.02
1.00
1.1
1.0
0.9
0.8
0.7
0.6
0.5
0.4
0.3
0
4k
8k
12k
16k
20k

Figure 4

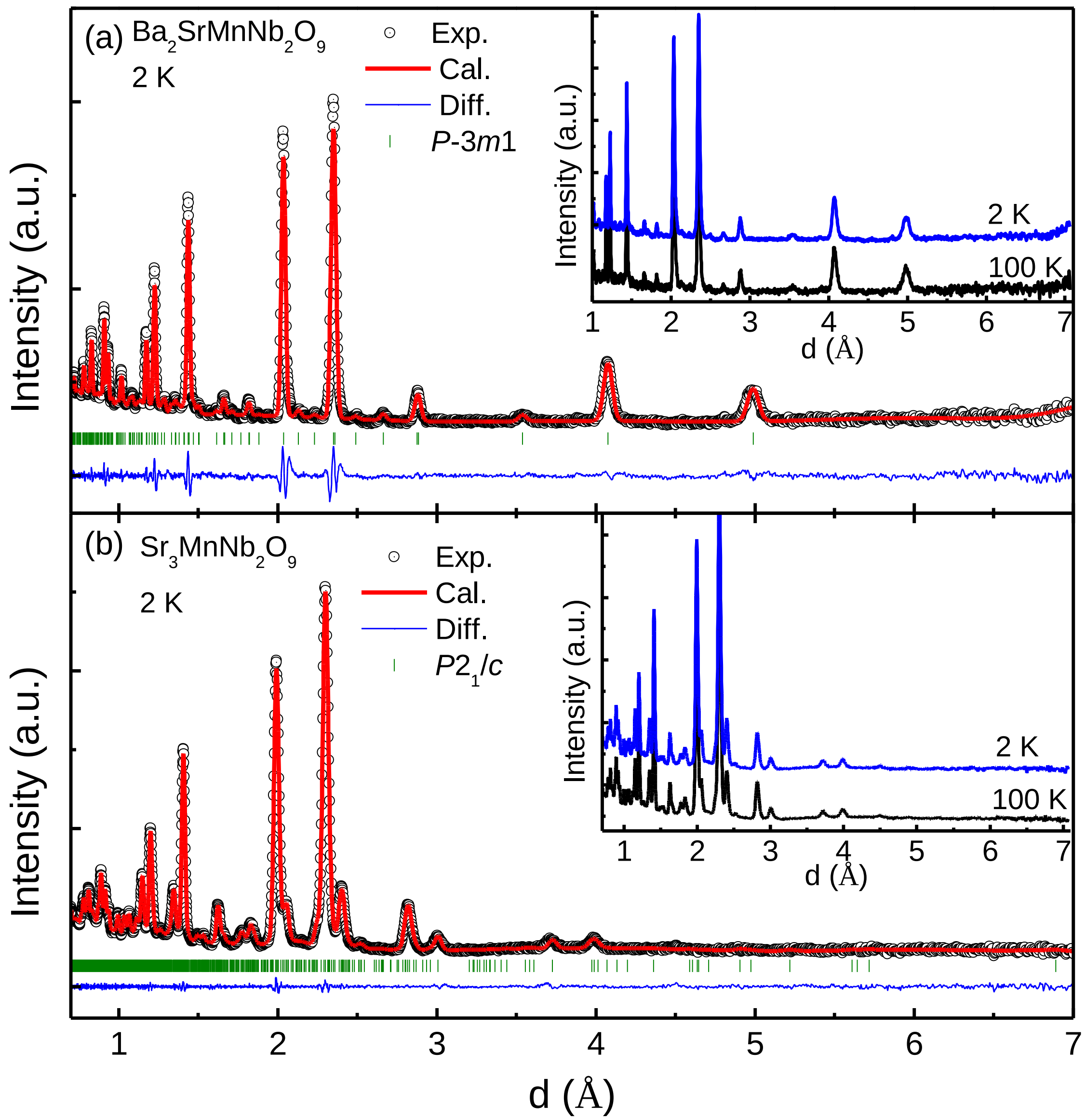

Figure 5

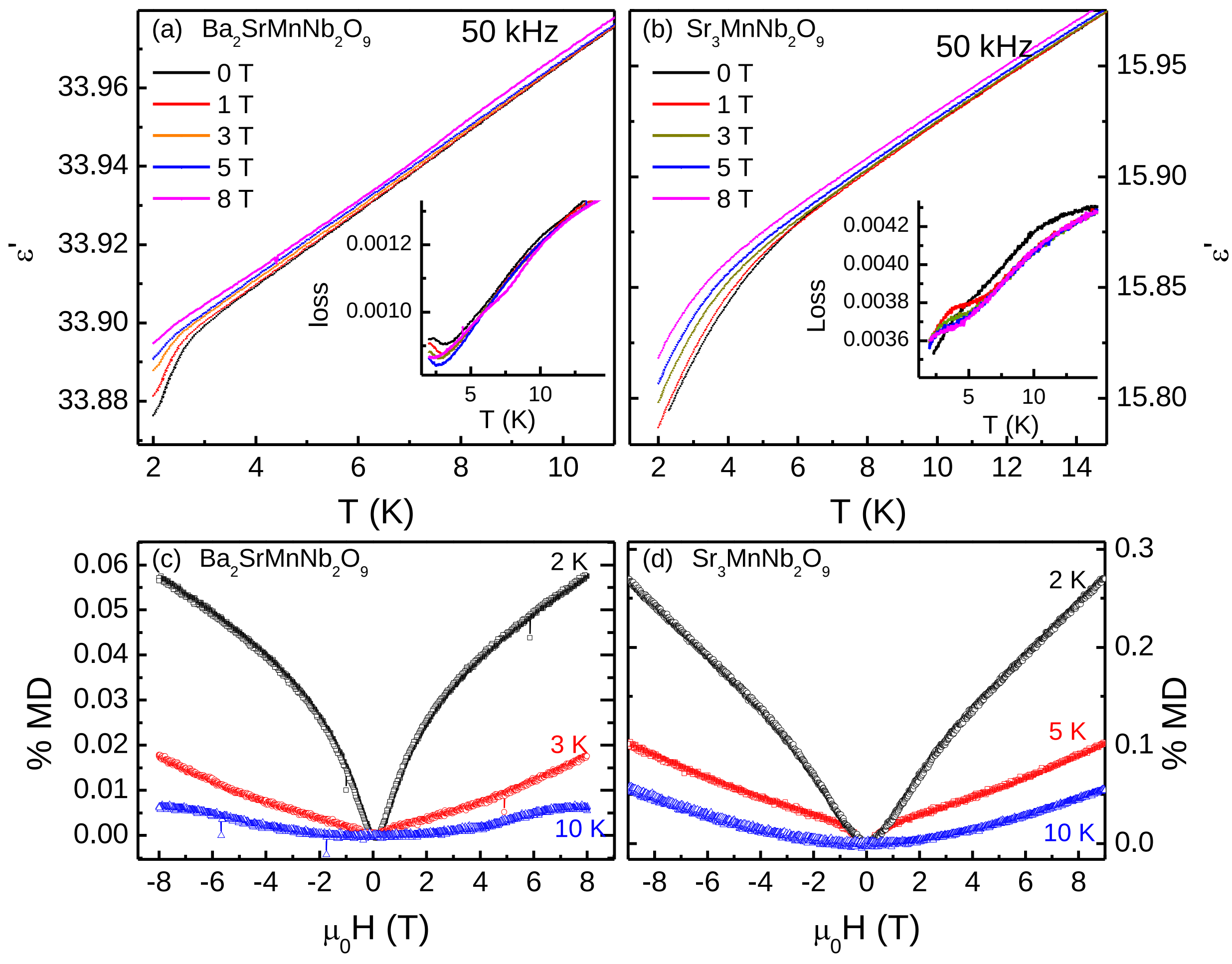

Figure 6